# Femtosecond dynamics of a polariton bosonic cascade at room temperature

**Author**
Fei Chen,[1] Hang Zhou,[2] Hui Li,[1*] Song Luo,[2] Zheng Sun,[1] Zhe Zhang,[2] Fenghao Sun,[1] Beier Zhou,[3] Hongxing Dong,[3] Huailiang Xu,[1] Hongxing Xu,[1] Alexey Kavokin,[4,5] Zhanghai Chen,[2*] and Jian Wu[1,6,7*]

**Affiliations**
[1]*State Key Laboratory of Precision Spectroscopy, East China Normal University, Shanghai 200241, China.*
[2]*Department of Physics, College of Physical Science and Technology, Xiamen University, Xiamen 361005, China.*
[3]*Shanghai Institute of Optics and Fine Mechanics, Chinese Academy of Sciences, Shanghai 201800, China.*
[4]*School of Science, Westlake University, Zhejiang 310024, China*
[5]*Institute of Natural Sciences,Westlake Institute for Advanced Study, Zhejiang 310024, China*
[6]*Collaborative Innovation Center of Extreme Optics, Shanxi University, Taiyuan, Shanxi 030006, China.*
[7]*CAS Center for Excellence in Ultra-intense Laser Science, Shanghai 201800, China.*

**Abstract**

Whispering gallery modes in a microwire are characterized by a nearly equidistant energy spectrum. In the strong exciton-photon coupling regime, this system represents a bosonic cascade: a ladder of discrete energy levels that sustains stimulated transitions between neighboring steps. In this work, by using femtosecond angle-resolved spectroscopic imaging technique, the ultrafast dynamics of polaritons in a bosonic cascade based on a one-dimensional ZnO whispering gallery microcavity is explicitly visualized. Clear ladder-form build-up process from higher to lower energy branches of the polariton condensates are observed, which are well reproduced by modeling using rate equations. Moreover, the polariton parametric scattering dynamics are distinguished on a timescale of hundreds of femtoseconds. Our understanding of the femtosecond condensation and scattering dynamics paves the way towards ultrafast coherent control of polaritons at room temperature, which will make it promising for high-speed all-optical integrated applications.

## MAIN TEXT

### Introduction

The strong coupling between excitons and cavity confined photons can lead to formation of hybrid half-light half-matter bosonic quasiparticles, i.e. the exciton-polaritons (EPs) (*1,2*). EPs bridge the gap between atomic physics and condensed matter physics by sharing many common characteristics with coupled atomic states. Phenomena such as Rabi splitting (*3*) and Bose-Einstein Condensation (BEC) in atomic systems can be found in the solid-state counterparts in a non-equilibrium regime (*4-6*). When the interparticle distances decrease to the level of their de Broglie wavelengths, polaritons can accumulate massively on a single quantum state at high temperatures due to their extremely light effective masses (*7*). Polariton condensates are of a unique driven-dissipative nature compared to conventional atomic BEC. They attract a significant interest because of the recently discovered macroscopic coherent phenomena including superfluidity (*8*), vortex formation (*9-11*), soliton propagations (*12,13*). Importantly, the excitonic constituent introduces strong coupling between polaritons through Coulomb interactions. This gives rise to a high nonlinearity in the interactions among polariton ensembles. Processes such as parametric scattering and bosonic amplification belong to this category (*14-17*). In addition to the profound fundamental physics involved in polariton BEC, the condensates show excellent performances as low-cost light source elements (*18-23*), ultrafast transistors and switches (*24-26*), which are all central building blocks of polaritonic integrated devices.

Optically confined semiconductor structures such as microcavities, micropillars, microwires etc., have an excellent potential for photonic band engineering. In the strong coupling regime, they can efficiently control the density of states of bosonic quasiparticles, i.e. the EPs. Theoretical proposals based on specific designs of microsystems sustaining polariton modes include Berry phase interferometers (*27*), polariton neurons (*28*), polariton qubits (*29*), bosonic cascades (*30*) etc. Here we focus on the experimental realization of bosonic cascades, where stimulated transitions in a ladder of equidistant energy levels are expected to result in the generation of coherent radiation (*31*) and non-classical light (*32*). The relaxation processes in a bosonic cascade can be enhanced by bosonic stimulation thus giving rise to an extremely high quantum efficiency without the need to realize population inversion. They intrinsically differ from the fermionic lasers. Attempts to the experimental realization of these effects with use of parabolic quantum wells embedded in planar microcavities (*33*) were not crowned by a decisive success so far. Here we demonstrate a room temperature bosonic cascade based on a one-dimensional (1D) ZnO microwire. In our experiments, polariton condensates occupy subsequently four neighboring quantum states in a quasi-equidistant energy spectrum. We visualize the femtosecond dynamics of stimulated transitions of exciton-polaritons between the condensates formed by whispering gallery modes of a ZnO microwire in the strong exciton-photon coupling regime. We argue that the ultrafast dynamics (*34-41*) can truly reveal the physics of bosonic cascades that has not been fully understood till now. The real time evolution of polariton condensates is characterized by multiple degrees of freedom involving energy, space and momentum. The processes of formation, relaxation as well as degenerate parametric scatterings of polariton condensates in a bosonic cascade are reported with an unprecedented time resolution. Our findings make a step forward in understanding of dynamics of many-body bosonic systems and offer the possibility to realize coherent manipulations, which may also promote the application of integrated polaritonic devices operating at room temperature.

## Results

**Polariton condensation cascade**
Polaritons excited by moderate intense laser fields in a 1D ZnO whispering gallery microcavity (WGM) usually involve condensation on multiple branches accompanying complicated parametric scattering processes. The photoluminescence (PL) measurements for the static states typically show a very complex pattern (presented in Fig. 1(a)) in which the superposed processes are hard to be disentangled. Using our femtosecond angle-resolved spectroscopic imaging (FARSI) technique, the dynamics of polariton condensation on multiple lower polariton (LP) branches, as well as the complex parametric scattering processes, can be clearly visualized. In our work, the PL distribution as a function of energy and momentum is recorded via the TE polarization which dominates the whole spectra when the 1D ZnO WGM is excited nonresonantly. The underlying dynamics of polariton condensation formed in 1D ZnO WGM is schematically shown in Fig. 1(b). At the leading edge of the femtosecond excitation pulse, exciton reservoir can be ignited almost instantly at femtosecond time scales. The density of the exciton reservoir becomes larger and larger as the field strength of the pumping pulse increases. The repulsive interactions between excitons become efficient and the excitons can be scattered to the surroundings. As the exciton reservoir relaxes, the screening effect on the cavity polaritonic system changes gradually, thus will result in a time-dependent shift of the detuning between exciton and cavity photon modes, which eventually affects the polariton condensation dynamics (*42*). Early study based on the ZnO microwire has also shown that the competition between media loss and cavity loss will give rise to an optimized whispering gallery mode with the highest quality factor (Q-factor) (*43*), polariton condensates are preferentially formed on such optimized mode. The exciton reservoir holding a large population can be strongly coupled to a cavity mode exhibiting high Q-factor to form polaritons on a certain LP branch, i.e., along Pathway 1 presented in Fig. 1. As the exciton reservoir relaxes, polaritons on the next LP branch with lower energy can be built from the exciton reservoir through Pathway 2. These polaritons experience evaporative cooling through polariton-phonon, polariton-polariton interactions and will gradually decay down to the ground state (*44*). When a threshold density is exceeded, a huge number of polaritons will massively accumulate onto one single quantum state, thus forming a condensate. It is theoretically predicted that the Bosonic ensemble on these nearly equidistant LP branches can transit through the energy ladders, generating terahertz (THz) radiation (*31*), as indicated by the yellow dashed arrow in Fig. 1(b). Such Bosonic cascading processes could dramatically influence the whole dynamics of the polariton condensation in a 1D ZnO microcavity. A ladder-form polariton condensation from higher to lower energy LP branches is observed that constitutes an unambiguous evidence for the bosonic cascade effect. Here we report the first observation of the femtosecond dynamics of polariton condensation formed on multiple quasi-equidistant states in a bosonic cascade at room temperature.

The dispersion mapping of the ZnO microwire is presented in Fig. 2(a) with multiple LP branches (marked by $P_{1-4}$) obtained from the plane wave model (*45*). Polariton condensation formation and relaxation on the four distinct LP branches are recorded with a time resolution of 30 fs for an excitation strength of $5P_{th}$ ($P_{th}$ is the threshold power for condensation formation). The time-integrated PL distribution as a function of energy and momentum (shown in Fig. 2(b)) exhibits a complex pattern. In the time-resolved measurement, a ladder-form build-up of the PL distributions from polariton condensation starting from the highest LP branch $P_4$, and successively transits towards the lower energy branches, i.e., $P_{3-1}$, are shown explicitly in Fig. 2 (c). The time delays between the adjacent branches are found to be around hundreds of femtoseconds ($\tau_{43}$: -100 fs, $\tau_{32}$: 310 fs, $\tau_{21}$: 110 fs. Here the $\tau_{ij}$ represents the time interval between the maximum population for the i-th and the j-th LP branch). The here-observed underlying dynamics is confirmed by the modeling using rate equations (*46,47*). The calculation results are compared with the experimental data and are shown in Fig. 2 (d-f). The detuning effect and bosonic cascading effect are both explicitly included in the equations. As shown in Fig. 2 (d), the nice agreement has been achieved between the simulation results and the experimental data. In our simulation, the detuning-dependent effect is considered by fitting the Q-factor for different energies at various LP branches. From that we can obtain empirical decay rates for the four LP branches (*43*). Optimization of all the other fitting parameters are achieved by using the genetic algorithm. As can been seen in Figs. 2 (d-f), only by including the coupling between adjacent LP branches can the simulation results reach a reasonable agreement with the measured data, which serve as a solid proof for the existence of cascading dynamics. The observed very first build-up of polariton condensate on the $P_3$ branch can be originate from the fact that the $P_3$ branch lie around the optimized wavelength range with the maximum Q-factor. The Q-factors for the other branches with shorter or longer wavelengths decrease rapidly. A slight deviation for the $P_1$ branch approaching the end of the multi-branch build-up process might be from the over-simplified approximation in our theoretical model, or from the imperfection of the ZnO microrod. Theoretically, such dynamics of polariton condensates in a bosonic cascade can give rise to THz radiation, which is not measured in the present work. As a matter of fact, the decay of the exciton reservoir is also accompanied by energy relaxation, which cannot be explicitly described by our theoretical model, but is shown in the time-dependent measurements (Fig. 2 (c)). The resulting polariton dynamics undergoes redshift with different characteristic times. This can be attributed to different relaxation mechanisms of the polariton condensates. In our measurements using ultrafast injection, the blueshift due to the polariton-polariton interaction cannot be distinguished because it happens so quickly that it is out of the scope of our observation. The data we reported here corresponds to the time domain after the blueshift. Each polariton condensate formed on the $P_{1-4}$ branches lasts for 560-860 fs. It indicates that each branch outputs a polariton lasing with femtosecond pulse duration.

When the pump power goes down, the transient injection to the exciton reservoir thus the polariton condensation dynamics can be tailored. Similar ladder-form build-up processes of the polariton condensation are also obtained on the P$_{3,4}$ branches at an excitation strength of 2P$_{th}$. As shown in Fig. 3, the overall dynamics are in comparison to the observations taken under higher excitation power. However, it shows a slower build-up and decay behavior. The coupling strength between the involved LP branches and thus the whole dynamics can be tailored by tuning the instantaneous injection populations. Based on these, the dynamics of the cascade of polariton condensates at room temperature can be controlled with femtosecond temporal resolutions through manipulating the excitation waveform.

**Polariton parametric scattering**
Polariton parametric scatterings (PPS) can induce entangled signal and idler ensembles which is why it is highly important for applications in quantum information processing (*17,48,49*). The polariton condensates in 1D microcavity are ideal sources for parametric scattering, due to the existence of rich inter-branch scattering channels and easy implementation of the phase matching condition. Complex scattering processes in 1D ZnO WGM have been predicted theoretically (*50*). However, due to the harsh requirement for time resolution, the sub-picosecond scattering dynamics has not been resolved experimentally. Here we boost up the time resolution down to femtoseconds and explicitly show the ultrafast dynamics of the PPS processes in 1D ZnO WGM at room temperature.

In Fig. 4, the time-resolved momentum distributions of the polariton condensates are illustrated for the P$_{2-4}$ branches, respectively. The scattered signal and idler pairs (labeled with S$_{L/R}$) at both sides of the condensate with degenerate energy and opposite momentum can be recognized. After an ultrafast injection, the formation of polariton condensation close to the ground state takes hundreds of femtoseconds to arrive at the maximum population. During this time, the momentum distributions of the condensate become broadened as the polariton-polariton interactions are enhanced (as shown in Figs. 4 (a-c)). After reaching the peak population, the momentum distribution of the polaritons in each LP branch experiences a clear shrinking. Both the build-up and the relaxation are accompanied by a time-dependent redshift observed in the energy domain (shown in Fig. 2 (c)). In general, polariton condensates are formed firstly at the ground states of a given LP branch, and part of the polaritons can be scattered parametrically towards the adjacent LP branches after hundreds of femtoseconds, as schematically illustrated in Fig. 1(b). The lagging time can be clearly illustrated by projecting the signals in Figs. 4 (a-c) onto the time axis. As shown in Fig. 4 (d), the magenta and blue curves represent the scattered signals to emission angles around ±36.3° from the P$_2$ branch. The peak positions of the scattered signals are delayed with respect to the source emitted at around 0° (presented by the yellow dashed curve) by about 290 fs. Comparable features of the scattering pairs can be recognized for another LP branch P$_4$ as shown in Fig. 4(g), where the scattering happens in around 230 fs. Particularly, the parametric scattering towards the next two neighboring branches are observed for the huge polariton condensate initially formed on the P$_3$ branch. A longer delay of about 490 fs can be found for the scattering pairs observed at around ±42.7° (shown in Fig. 4 (f)), compared to the 330fs delay time obtained for the scattering pairs at around ±34.7° (shown in Fig. 4 (e)). The different mechanisms occurring in different time scales can be distinguished by co-analyzing the energy and momentum distributions of the PL emission. We found that the slower redshift happened beyond, e.g. 3 ps for the P$_3$ LP branch shown in Fig. 2 (c) can be mainly attributed to the PPS process.

Degenerate inter-branch parametric scatterings can also be observed for the low excitation cases. At the pump power of 2P$_{th}$ (as shown in Fig. 3 (b)), clear scattering channels can be distinguished at emission angles around ±34.2°. The projected data in Fig. 3 (d) show that the scattering signal towards the adjacent mode appears with a lagging time of about 380 fs. The time for the polariton condensates to be scattered from the ground state of P$_3$ to the adjacent LP branch at the excitation of 2P$_{th}$ is comparable to the 330 fs scattering time obtained under stronger excitation condition. As shown in Fig. 4 (b), a larger polariton population on the P$_3$ branch can be formed, and the PPS may take place towards further LP branches at larger angles due to the stronger interactions in the polariton ensemble.

**Discussion**
In this work, we have visualized the femtosecond dynamics of polariton condensation in a bosonic cascade at room temperature in a 1D ZnO WGM using the FARSI technique. The cascading build-up process of the polariton condensation from higher to lower energy LP branches is documented for various excitation powers. The characteristic build-up and relaxation times of polariton condensation on each of the steps of the cascade are obtained with femtosecond resolutions which can be precisely tailored by the optical excitation conditions. Theoretical simulations utilizing rate equations have shown that the bosonic cascading effect plays an inevitable role in the whole dynamics. Moreover, the ultrafast dynamics of rich parametric scattering channels are revealed, where the scattered signals towards the adjacent LP branches are blooming with a lagging time of hundreds of femtoseconds with respect to their scattering sources. This dynamics is original and goes beyond the theoretical expectations for the bosonic cascade dynamics that did not consider subbands associated with every step of the ladder (*30-33*). The highly efficient parametric scattering channels discovered here can be applied potentially in quantum information processing. Our work unveils the details of femtosecond dynamics of polariton condensation in an extreme time domain. Further work needs to be done in order to evidence the THz lasing and non-classical light generation in ZnO microwires.

**Materials and Methods**
FARSI technique is developed to explore the ultrafast dynamics of polariton condensation in 1D microcavity systems at room temperature. In our experiment, femtosecond pulses at central wavelength of 350 nm were used to excite the ZnO microcavity at room temperature. The duration of the excitation pulses was estimated to be around 400 fs at the interaction site. Angle-resolved measurements were performed on the Fourier plane by using an Andor spectrometer (SR-500i-B1, equipped with an intensified Charge Coupled Device (ICCD), iStar). In addition, an ultrafast switch based on optical Kerr effect [51] provided a platform to realize the femtosecond-resolved measurement. With such ultrafast shutter, the PL in both the energy and momentum degrees of freedom can be visualized frame by frame with high time resolutions. The dispersion arising from all the related optical elements between the ZnO microrod and the Kerr medium has been carefully characterized. In the present setup, the time resolution of the ultrafast gating is estimated to be around 30 fs through highly nonlinear interactions.

The 1D ZnO microrods are synthesized by chemical vapor deposition (CVD) method which retain hexagonal cross sections with a uniform radius of about 1.8 $\mu$m and a length of about hundred microns. The high quality ZnO WGM possesses a Q factor over 1000. The large exciton binding energy of about 60 meV makes ZnO microcavity an ideal candidate for manipulating polariton condensation at room temperature. In our work, polariton condensates onto preferred LP branches near $k_{//}=0$ when the excitation power exceeds the threshold ($P_{th}$ = 450 nW for the femtosecond excitation pulses at central wavelength of 350 nm). Both static state angle-resolved PL spectroscopy as well as the femtosecond-resolved measurements are carried out at room temperature for different excitation strengths.

We have used the rate equations characterizing multiple LP branches of the cascade (*47,48*) to model the dynamics of the polariton condensation under different excitation powers. The rate equations are written as follows.

$$\frac{\partial n_R}{\partial t} = P(t) - \gamma_R n_R - \sum_{i=1}^{4} R_i n_R (n_i + 1) \tag{1}$$

$$\frac{\partial n_4}{\partial t} = -\gamma_4 n_4 + R_4 n_R (n_4 + 1) - R_{43} n_R (n_3 + 1) \tag{2}$$

$$\frac{\partial n_k}{\partial t} = -\gamma_k n_k + R_k n_R (n_k + 1) + R_{(k+1)k} n_{k+1} (n_k + 1) - R_{k(k-1)} n_k (n_{k-1} + 1) \text{ (k=2,3)} \tag{3}$$

$$\frac{\partial n_1}{\partial t} = -\gamma_1 n_1 + R_1 n_R (n_1 + 1) + R_{21} n_2 (n_1 + 1) \tag{4}$$

Where P(t) is the pump power. $n_R$ and $\gamma_R$ represent the density and the decay rate of the exciton reservoir, respectively. $n_k$ (k=1-4) represents the polariton density on the LP branches of $P_{1-4}$, $\gamma_k$ is the corresponding decay rate. $R_k$ (k=1-4) stands for the scattering rate from the exciton reservoir to the k-th LP branch $P_k$. $R_{ij}$ represents the inter-branch scattering rate between the adjacent LP branches, i.e., indicating the Bosonic cascading effect. Former works have shown that for such 1D WGM system, the competition between cavity loss and media loss can give rise to an optimized cavity mode with the highest Q factor. The Q factors for different LP branches at the quasi-equidistant energies are obtained to decide the decay rates on each LP branch. All the other parameters in Eqns. 1-4 are optimized through genetic algorithm.

**Acknowledgments**

**General**: We thank Dr. Z. Wang for the fruitful discussions.

**Funding:** This work is supported by the National Key R&D Program of China (Grant nos. 2018YFA0306303, 2018YFA0306304); the National Natural Science Fund (Grants No. 92050105, 91950201, 61690224, 11761141004, 11834004, 11621404, 11704124, 11674069 and 11574205); the Project supported by the Shanghai Committee of Science and Technology, China (Grant No. 19ZR1473900). AK acknowledges Project No. 041020100118 supported by Westlake University and Programme 2018R01002 funded by the Leading Innovative and Entrepreneur Team Introduction Programme of Zhejiang.

**Author contributions:** J. W., Z. C. and H. L. conceived the idea and initiated the study. F. C., H. L., S. L., Z. Z., and F. S. conducted the experimental work. H. Z. and F. C. performed the simulations. All authors contributed to the data analysis and writing the manuscript. J. W., Z. C. and H. L. supervised and guided the work.

**Competing interests:** Include any financial interests of the authors that could be perceived as being a conflict of interest. Also include any awarded or filed patents pertaining to the results presented in the paper. If there are no competing interests, please state so.

**Data and materials availability:** Data and materials may be requested from the authors.

**Figures**

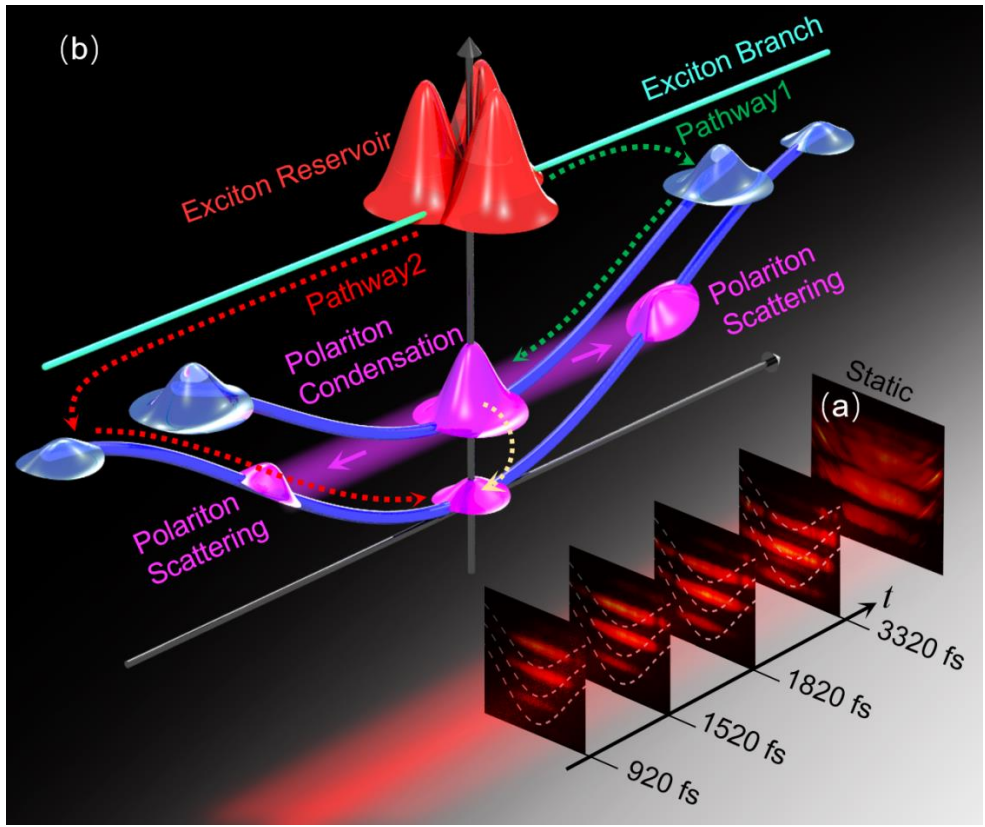

**Fig. 1. Schematic illustration of polariton dynamics in a Bosonic cascade.** (a) The schematic drawing of the polariton condensation and parametric scattering dynamics. (b) The femtosecond angle-resolved spectroscopic imaging (FARSI) measurement.

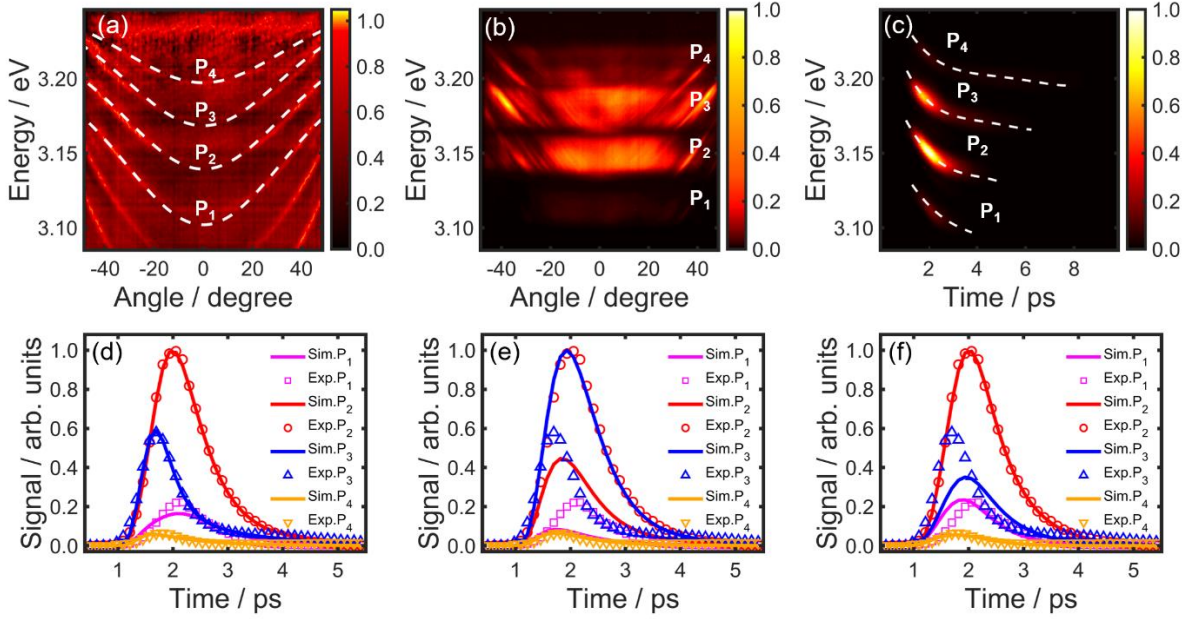

**Fig. 2. Polariton condensation dynamics in an equidistant energy spectrum.** (a) The dispersion mapping of the ZnO WGM obtained at an excitation power below the condensation threshold. Dashed curves are the fitting results for the LP branches labeled as $P_{1-4}$, respectively. (b) Time-integrated 2D imaging of the photoluminescence (PL) distributions as a function of emission angle and energy obtained in the k-space with the pump power at $P = 5P_{th}$. (c) Time-dependent PL distribution in the energy degree of freedom for the four branches. They all show the condensation relaxation with a redshift as the time elapses. The dashed curve helps to guide the eyes. (d-f) The calculation and experimental results for time-dependent populations of the polariton condensates for the four LP branches. The solid line/scattered dots are for describing the populations of the polariton condensates obtained from simulation/measured results in the $P_{1-4}$ branches. Here, (d) shows the fitting results by involving bosonic cascading effect in the rate equations, (e) shows the simulations results by turning off the cascading effect in the results of (d). (f) shows the fitting results using rate equations without cascading effect. By comparing the data shown in (d-f) we can see that the cascading effect plays a decisive role in the underlying dynamics.

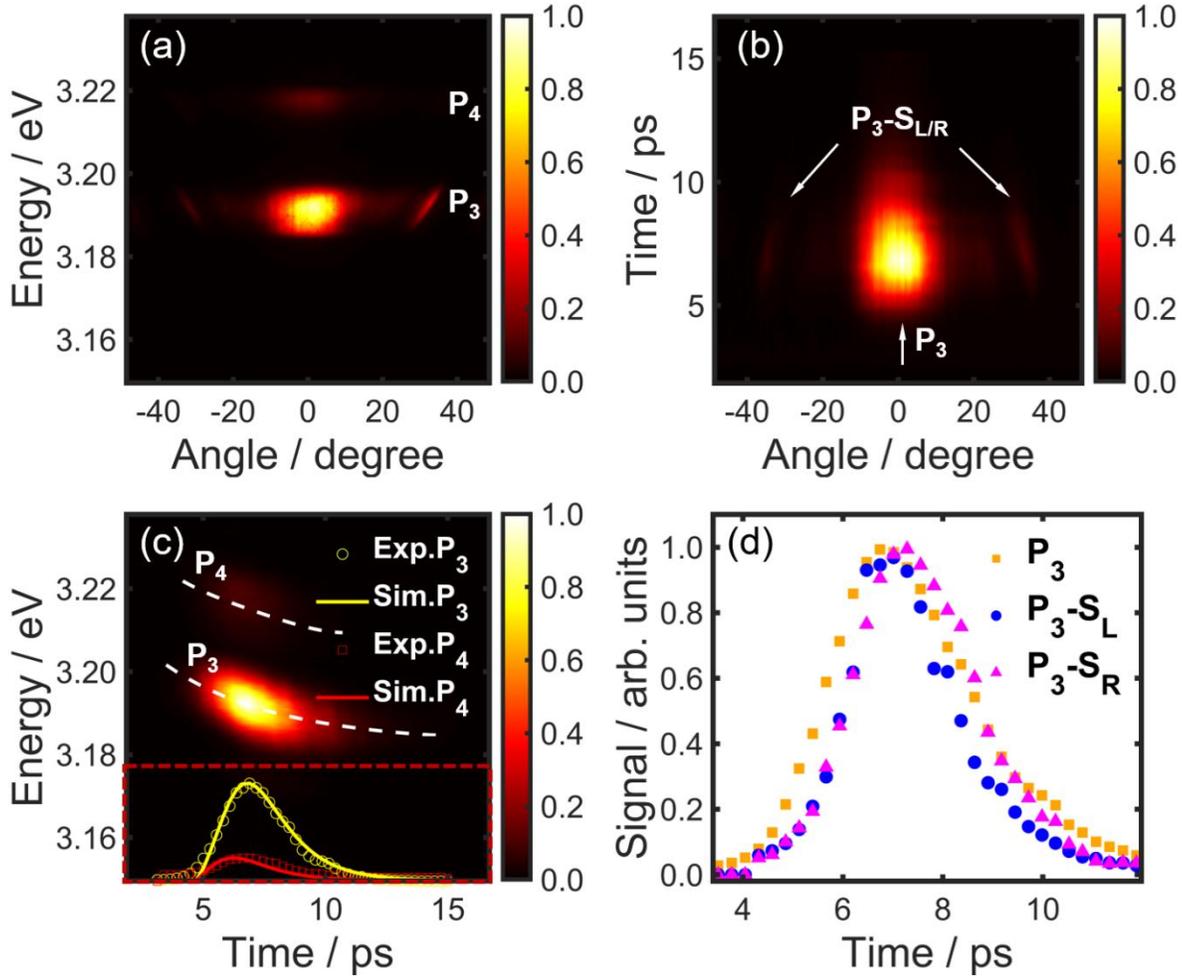

**Fig. 3. The static and femtosecond-resolved polariton condensation dynamics at lower pump power (P = 2P$_{th}$).** (a) Time-integrated 2D PL distributions as a function of energy and emission angle in the momentum space. (b) PL distributions as a function of angle and time in the momentum space, extracted for the P$_3$ LP branch. (c) PL emission as a function of energy and time. The dashed curve helps to guide the eyes which indicates the redshift during polariton condensation relaxation. The measured (scattered) and calculated (solid curve) yields of polariton condensate on the P$_3$ and P$_4$ LP branches are plotted by the yellow and red curves at the lower part. (d) The corresponding time-dependent polariton populations by projecting the measured data in (b) onto the time axis for the indicated scattering source (labeled as P$_3$) and scattering signals (labeled as P$_3$-S$_{L/R}$ in (b)).

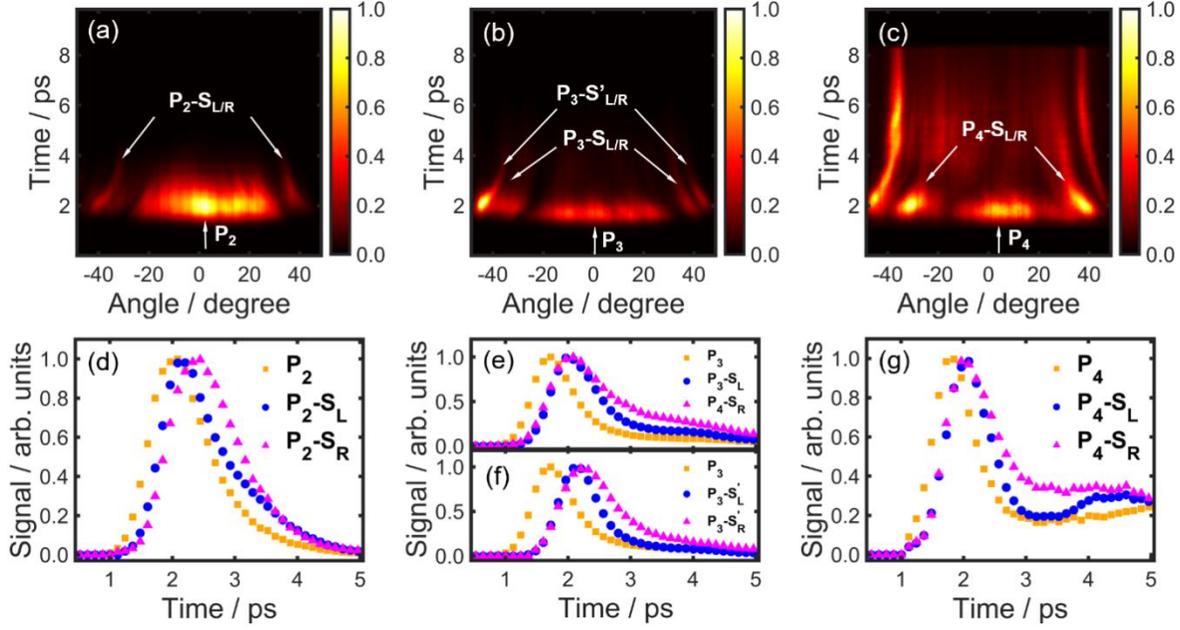

**Fig. 4. Polariton parametric scattering dynamics.** (a-c) PL as a function of emission angle and time extracted for the $P_{2,3,4}$ branches in the momentum space, respectively. (d) The corresponding time-resolved signal by projecting the data shown in (a) onto the time axis for the selected regions indicated in (a). The dynamics of the scattering source close to the center (labeled as $P_2$) and the scattered parts (labeled as $P_2$-$S_{L/R}$) are plotted separately, showing characteristic delays. (e, f, g) Similar to (d) but for polariton condensates on the $P_3$ and $P_4$ branches, respectively.